\documentclass{article}
\usepackage{setspace}
\usepackage[utf8]{inputenc}
\usepackage{amsfonts,amsmath,amssymb,mathtools}
\usepackage[left=2cm,right=2cm,top=2cm,bottom=2cm,bindingoffset=0cm]{geometry}
\usepackage{graphicx}
\usepackage{authblk}
\usepackage{xcolor}
\usepackage{hyperref}
\usepackage{authblk}
\usepackage{slashed}
\usepackage{braket}
\usepackage{comment}
\usepackage{pb-diagram}
\usepackage{wasysym}
\usepackage{amsthm}
\usepackage{cancel}
\usepackage{tikz}\usetikzlibrary{tikzmark}
\usetikzlibrary{fadings} 
\usetikzlibrary{positioning}
\usetikzlibrary{backgrounds} 
\usetikzlibrary{decorations.pathreplacing}
\usetikzlibrary{arrows}
\definecolor{airforceblue}{rgb}{0.36, 0.54, 0.66}	
\definecolor{beige}{rgb}{0.96, 0.96, 0.86}
\definecolor{bittersweet}{rgb}{1.0, 0.44, 0.37}
\definecolor{melon}{rgb}{0.99, 0.74, 0.71}
\definecolor{mustard}{rgb}{1.0, 0.86, 0.35}
\definecolor{lava}{rgb}{0.81, 0.06, 0.13}
\definecolor{magnolia}{rgb}{0.97, 0.96, 1.0}
\definecolor{lavendermist}{rgb}{0.9, 0.9, 0.98}
\definecolor{lavendergray}{rgb}{0.77, 0.76, 0.82}
\definecolor{palepink}{rgb}{0.98, 0.85, 0.87}
\definecolor{palesilver}{rgb}{0.79, 0.75, 0.73}
\definecolor{cadetgrey}{rgb}{0.57, 0.64, 0.69}
\definecolor{anti-flashwhite}{rgb}{0.95, 0.95, 0.96}
\colorlet{Light0anti-flashwhite}{anti-flashwhite!70!white}
\colorlet{Lightanti-flashwhite}{anti-flashwhite!50!white}
\colorlet{Light2anti-flashwhite}{anti-flashwhite!30!white}
\definecolor{linkcolor}{rgb}{0,0,1}
\definecolor{urlcolor}{rgb}{0,0,1}

\usepackage{tikz-cd}
\usetikzlibrary{decorations.markings}
\usetikzlibrary{positioning}
\usepackage{braids}
\usepackage{array}
\usepackage{ytableau}
\usepackage{longtable}
\usepackage{empheq}
\usepackage{multicol}
\usepackage{mathrsfs}
\usepackage{diagbox}
\usepackage[vcentermath]{youngtab}
\usepackage[natbib=true, backend = bibtex, style=numeric-comp, sorting=none,maxbibnames=99]{biblatex}
\hypersetup{pdfstartview=FitH,  linkcolor=linkcolor,urlcolor=urlcolor,citecolor=urlcolor, colorlinks=true}

\newcommand\bem{\begin{pmatrix}}
\newcommand\eem{\end{pmatrix}}
\newcommand\beq{\begin{equation}}
\newcommand\eeq{\end{equation}}
\newcommand\beqs{\begin{equation*}}
\newcommand\eeqs{\end{equation*}}

\date{}

\def\be{\begin{eqnarray}}
\def\ee{\end{eqnarray}}

\definecolor{red}{rgb}{1,0,0}
\definecolor{orange}{rgb}{1,0.5,0}
\definecolor{violet}{rgb}{0.7,0,1}

\begin{document}

\title{\bf Tug-the-hook symmetry for quantum 6j-symbols
}
\author[1,3]{{\bf E.~Lanina}\thanks{\href{mailto:lanina.en@phystech.edu}{lanina.en@phystech.edu}}}
\author[1,2,3]{{\bf A. Sleptsov}\thanks{\href{mailto:sleptsov@itep.ru}{sleptsov@itep.ru}}}

\vspace{5cm}

\affil[1]{Moscow Institute of Physics and Technology, 141700, Dolgoprudny, Russia}
\affil[2]{Institute for Information Transmission Problems, 127051, Moscow, Russia}
\affil[3]{NRC "Kurchatov Institute", 123182, Moscow, Russia\footnote{former Institute for Theoretical and Experimental Physics, 117218, Moscow, Russia}}
\renewcommand\Affilfont{\itshape\small}

\maketitle

\vspace{-7cm}

\begin{center}
	\hfill MIPT/TH-19/22\\
	\hfill ITEP/TH-22/22\\
	\hfill IITP/TH-21/22
\end{center}

\vspace{4.5cm}

\begin{abstract}

{
We introduce a novel symmetry for quantum 6j-symbols, which we call the tug-the-hook symmetry. Unlike other known symmetries, it is applicable for any representations, including ones with multiplicities. We provide several evidences in favour of the tug-the-hook symmetry. First, this symmetry follows from the eigenvalue conjecture. Second, it is shown by several new examples of explicit coincidence of 6j-symbols with multiplicities. Third, the tug-the-hook symmetry for Wilson loops for knots in the 3d Chern-Simons theory implies the tug-the-hook symmetry for quantum 6j-symbols. An important implication of the analysis is the generalization of the tug-the-hook symmetry for the Chern-Simons Wilson loops to the case of links.
}
\end{abstract}
\setcounter{equation}{0}
\section{Introduction}

\textit{Racah coefficients} or \textit{6j-symbols} (see Subsection~\ref{Racah} for definitions) provide isomorphism between two different fusions in tensor product of three representations. Racah coefficients are common objects in theoretical and mathematical physics. They often occur in problems where tensor products of representations are taken. In order to effectively deal with such problems, one needs to deeply understand the internal structure of Racah coefficients and their analytical dependence on parameters. 

A basic example in physics is recoupling of three angular momenta in quantum mechanics~\cite{landau2013quantum}. There, Racah matrix provides transformation between two bases corresponding to different fusions of momenta. There is a huge list of physical applications of 6j-symbols -- they appear in description of Landau-Pomeranchuk-Migdal effect in quark-gluon plasma~\cite{arnold2019landau}, in problems of nuclear spectroscopy~\cite{hecht1975simple,chang1971distribution}, in lattice gauge theory~\cite{aroca1998path}, in problems related to the quantum states of ultracold alkaline earth atoms~\cite{gorshkov2010two}, in semiconductor theory for qubits construction~\cite{durst2020quadrupolar}. More involved examples are modular transformations of conformal blocks~\cite{moore1989classical,kaul1992three,ramadevi2013n} and calculation of observables in the Chern-Simons theory by Reshetikhin-Turaev approach~\cite{Reshetikhin,guadagnini1990chern2,reshetikhin1990ribbon,turaev1990yang,reshetikhin1991invariants}. 

Let us stress in more detail the physical meaning of objects under study in conformal field theories. One of the main objects in CFT is a \textit{monodromy operator}, which turn around loops or replace (\textit{half-monodromy}) primary fields. In theory of integrable systems and in our research, these operators are called \textit{$\mathcal{R}$-matrices} and in WZW conformal model they are known as \textit{braiding operators}~\cite{moore1989classical,Gu_2015}. Usually, eigenvalues of such monodromy operators are known/computable. However, it is not enough if one studies correlation functions of more than two primary fields and one needs to know monodromy operators of all neighboring pairs of fields. Then, one should be able to switch between bases in which the corresponding monodromy operators are diagonal. In CFT, these transition matrices are called \textit{crossing} or \textit{fusion matrices} and correspond to quantum 6j-symbols, whose symmetries we study in this paper.

Racah coefficients are well-defined for finite-dimensional~\cite{racah1942theoryII,wigner2012group} and infinite-dimensional representations~\cite{groenevelt2006wilson,ponsot2001clebsch,ismagilov2006racah,derkachov20196j} of classical Lie groups, and for representations of quantum groups~\cite{kirillow1989representations}. In the latter case, Racah coefficients are called \textit{quantum}. In this paper, we consider only irreducible finite-dimensional representations of $U_q(\mathfrak{sl}_N)$ quantum algebra.

The eigenvalue hypothesis (see Subsection~\ref{EigenHyp}) tells that Racah coefficients are uniquely determined by normalized eigenvalues of the corresponding $\mathcal{R}$-matrices. A straightforward consequence is that two Racah coefficients are equal if $\mathcal{R}$-matrices eigenvalues coincide, or equivalently if $\mathcal{R}$-matrices differ by a multiplier and a particular unitary rotation. The hypothesis was suggested in~\cite{itoyama2013eigenvalue} for all equal incoming representations in tensor product and further generalized to the case when incoming representations can be different in~\cite{anokhina2014cabling}.

Most above mentioned physical examples involve irreducible finite-dimensional representations of classical Lie groups and algebras which correspond to the classical case $q=1$. However, the eigenvalue hypothesis can be formulated only for the quantum case. In the classical case $\mathcal{R}$-matrices are reduced to permutation matrices which permute two highest weight vectors and have unique eigenvalues equal to $\pm 1$. But it does not mean that corresponding Racah matrices for all representations are equal. Meanwhile, if for generic $q$ from the eigenvalue hypothesis, we obtain the equality of Racah matrices, this equality still holds in particular case of $q=1$. This fact allows one to get results for the classical case, which is of great importance in physics.

The eigenvalue conjecture has been proved for $U_q(\mathfrak{sl}_2)$ case in~\cite{Alekseev_2021}. Although there is no proof for the rank $N>2$, the hypothesis was checked in some cases. In the multiplicity-free case, there are exact expressions for Racah coefficients through $\mathcal{R}$-matrix eigenvalues in the case of all equal incoming representations for matrices of the size up to $5\times 5$~\cite{itoyama2013eigenvalue} and $6\times 6$~\cite{mironov2016universal} for three strands, and in the case when all incoming represenations can be different -- for matrices of the size up to $3\times 3$ for three strands~\cite{anokhina2014cabling} and for matrices of the size up to $5\times 5$ for four strands~\cite{Dhara_2018}. The situation becomes much more complicated when multiplicities occur, but even in this case when Racah matrices can be made block-diagonal, these blocks should satisfy the eigenvalue hypothesis~\cite{Bishler_2018}.

The eigenvalue hypothesis has lots of important applications. The one we use in our consideration is its ability to predict classes of symmetries of Racah coefficients~\cite{AMorozov_2018} and quantum knot invariants. However, explicit symmetries of 6j-symbols have been provided only for symmetric and their conjugate representations~\cite{Alekseev_2020,Alekseev_2021}. Another interesting implication is its connection~\cite{Mironov_2018} with the one-hook scaling property of the colored Alexander polynomial, which relates these polynomials for single-hook representations and the fundamental one. This symmetry unexpectedly connects quantum knot invariants with the Kadomtsev–Petviashvili integrable hierarchy~\cite{mishnyakov2021perturbative}.

In this paper, we introduce a new symmetry for quantum 6j-symbols (for other currently known symmetries see Subsection~\ref{Symm}). From the one hand, relying on the eigenvalue conjecture, we derive the tug-the-hook symmetry for Racah coefficients (see Section~\ref{TTH}). This symmetry is proved in the above mentioned cases where the eigenvalue hypothesis is confirmed (see also Subsection~\ref{EigHypProofs}). From the other hand, we present examples of coincidence of Racah matrices connected by the tug-the-hook transformation (see Subsection~\ref{TTHEx}). Moreover, the tug-the-hook symmetry has been proved for the colored HOMFLY polynomials (which are Wilson loops for the 3d Chern-Simons theory) for knots~\cite{Lanina:2021nfj}. Due to the fact that the HOMFLY polynomials are particularly constructed from Racah matrices, this symmetry should also hold for Racah coefficients (see Subsection~\ref{HOMFLY}). This independent evidence for the tug-the-hook symmetry for Racah coefficients also indirectly confirms the eigenvalue hypothesis. Another important implication we provide is the tug-the-hook symmetry for the HOMFLY polynomials for links (see Subsection~\ref{HOMFLY}).

\setcounter{equation}{0}
\section{Preliminaries}
In this section, we discuss basic definitions and properties of objects we use in our analysis -- quantum 6j-symbols (Subsection~\ref{Racah}), $\mathcal{R}$-matrices (Subsections~\ref{R-matrix} and~\ref{RviaRacah}) and the colored HOMFLY polynomials (Subsection~\ref{HOMFLYdefSec}). We also discuss the eigenvalue hypothesis and its origins in Subsection~\ref{EigenHyp}.

\subsection{Quantum Racah coefficients and 6j-symbols}\label{Racah}
In this subsection, we introduce \textit{quantum Racah coefficient} and its normalization -- \textit{6j-symbol}. In conformal field theories, Racah matrices are called \textit{crossing} or \textit{fusion matrices} and relate two types of conformal blocks coming from two different orders of operator product expansion in four point correlation function of primary fields.

Consider three finite-dimensional irreducible representations $V_{R_1}$, $V_{R_2}$, $V_{R_3}$ of the quantized universal enveloping algebra $U_q(\mathfrak{sl}_N)$. We assume that $q$ is a nonzero complex number which is not a root of unity, thus, all finite-dimensional representations are representations of highest weights, and they can be enumerated by Young diagrams. Recall that a Young diagram $\mu=\{\mu_1\geq\mu_2\geq\dots\geq\mu_l\}$ is constructed from highest weights $\{\omega_1,\omega_2,\dots,\omega_l\}$ of a representation $V_\mu$: $\mu_i=\sum_{k=i}^l\omega_k$ $\forall\,i=1,\dots,l$, and vice versa $\omega_i=\mu_i-\mu_{i+1}$. Since the tensor product $V_{R_1}\otimes V_{R_2}\otimes V_{R_3}$ is associative, there is a natural isomorphism:
\begin{equation}\label{Assoc}
    (V_{R_1}\otimes V_{R_2})\otimes V_{R_3}\rightarrow V_{R_1}\otimes(V_{R_2}\otimes V_{R_3})\,.
\end{equation}
One can expand tensor products of two representations into a direct sum of irreducible components:
\begin{equation}
 V_{R_1}\otimes V_{R_2}=\bigoplus_{R_{12}} \mathcal{M}_{R_{12}}^{R_1,R_2}\otimes V_{R_{12}}\,, \quad V_{R_2}\otimes V_{R_3}=\bigoplus_{R_{23}} \mathcal{M}_{R_{23}}^{R_2,R_3}\otimes V_{R_{23}}\,.
\end{equation}
Here $\mathcal{M}_{R_{12}}^{R_1,R_2}$ and $\mathcal{M}_{R_{23}}^{R_2,R_3}$ are the subspaces of highest weight vectors with highest weights corresponding to the Young diagrams $R_{12}\vdash|R_1|+|R_2|$ and $R_{23}\vdash|R_2|+|R_3|$ respectively. The dimensions of the spaces $\mathcal{M}_{R_{12}}^{R_1,R_2}$ and $\mathcal{M}_{R_{23}}^{R_2,R_3}$ are called \textit{multiplicities} of representations $V_{R_{12}}$ and $V_{R_{23}}$ respectively. 

Expand the second tensor products in~\eqref{Assoc}:
\begin{equation}
\begin{aligned}
 (V_{R_1}\otimes V_{R_2})\otimes V_{R_3}&=\bigoplus_{R_{12},R_{123}} \mathcal{M}_{R_{12}}^{R_1,R_2}\otimes\mathcal{M}_{R_{123}}^{R_{12},R_3}\otimes V_{R_{123}}\,, \\
 V_{R_1}\otimes(V_{R_2}\otimes V_{R_3})&=\bigoplus_{R_{23},R_{123}} \mathcal{M}_{R_{123}}^{R_1,R_{23}}\otimes\mathcal{M}_{R_{23}}^{R_2,R_3}\otimes V_{R_{123}}\,.
\end{aligned}
\end{equation}
Then, the associativity condition~\eqref{Assoc} implies the following definitions. \\
\textit{Racah matrix is the map}
\begin{equation}
    U\begin{bmatrix} R_1 & R_2 \\ R_3 & R_{123}
    \end{bmatrix}: \bigoplus_{R_{12}} \mathcal{M}_{R_{12}}^{R_1,R_2}\otimes\mathcal{M}_{R_{123}}^{R_{12},R_3}\rightarrow \bigoplus_{R_{23}} \mathcal{M}_{R_{123}}^{R_1,R_{23}}\otimes\mathcal{M}_{R_{23}}^{R_2,R_3}\,.
\end{equation}
\textit{Racah coefficients $U\left[\begin{array}{cc|c} R_1&R_2& R_{12} \\ R_3&R_{123} &R_{23}\end{array}\right]$ are elements of Racah matrix.}\\ \\
\textit{Wigner 6j-symbol is a normalized Racah coefficient:}
\begin{equation}
    \left\lbrace \begin{matrix}\label{6j_def}
	R_1 & R_2 & R_{12} \\
	R_3 & R_{123} & R_{23}
	\end{matrix} \right\rbrace =  \frac{1}{\sqrt{\text{qdim}(R_{12})\,\text{qdim}(R_{23})}}\, U\left[\begin{array}{cc|c} R_1&R_2& R_{12} \\ R_3&R_{123} &R_{23}\end{array}\right],
\end{equation}
\textit{where} qdim($R$) \textit{denotes the quantum dimension of representation $V_R$.}\\ \\
Explicit calculations of quantum Racah coefficients through highest weight vectors for different representations can be found in~\cite{Mironov_2015,AMironov_2016,AMironov2_2016,SS,Bai_2018}.

In what follows, we will utilize the following property~\cite{haase1985algebraic,haase1986symmetric,klimyk2012quantum}. \\ \\
\textit{Racah coefficients are invariant under the transposition of all representations:}
\begin{equation}\label{UT}
    U\left[\begin{array}{cc|c} R_1&R_2& R_{12} \\ R_3&R_{123} &R_{23}\end{array}\right]=U\left[\begin{array}{cc|c} R'_1&R'_2& R'_{12} \\ R'_3&R'_{123} &R'_{23}\end{array}\right],
\end{equation}
\textit{where $'$ denotes transposition of a Young diagram.}

\subsection{Currently known symmetries of quantum 6j-symbols}\label{Symm}
In this subsection we briefly review currently known symmetries of quantum 6j-symbols. \\
1. Quantum 6j-symbol is invariant under cyclic permutation of columns and under exchange of columns (with the corresponding conjugation of representations) up to a sign:
    \begin{equation}\label{permutation}
        \left\lbrace \begin{matrix}
	R_1 & R_2 & R_{3} \\
	r_1 & r_{2} & r_{3}
	\end{matrix} \right\rbrace = \left\lbrace \begin{matrix}
	R_2 & R_3 & R_{1} \\
	r_2 & r_{3} & r_{1}
	\end{matrix} \right\rbrace = \left\lbrace \begin{matrix}
	R_2 & R_1 & R_{3} \\
	\overline{r}_2 & \overline{r}_{1} & \overline{r}_{3}
	\end{matrix} \right\rbrace\,,
    \end{equation}
    where $\overline{R}$ is a representation conjugate to $R$. \\ \\
    2. If one exchanges elements in two neighboring columns
    \begin{equation}\label{exchange}
        \left\lbrace \begin{matrix}
	R_1 & R_2 & R_{3} \\
	r_1 & r_{2} & r_{3}
	\end{matrix} \right\rbrace = \left\lbrace \begin{matrix}
	\overline{R}_1 & r_2 & \overline{r}_{3} \\
	\overline{r}_1 & R_{2} & \overline{R}_{3}
	\end{matrix} \right\rbrace = \left\lbrace \begin{matrix}
	\overline{r}_1 & \overline{R}_2 & r_{3} \\
	\overline{R}_1 & \overline{r}_{2} & R_{3}
	\end{matrix} \right\rbrace = \left\lbrace \begin{matrix}
	r_1 & \overline{r}_2 & \overline{R}_{3} \\
	R_1 & \overline{R}_{2} & \overline{r}_{3}
	\end{matrix} \right\rbrace\,.
    \end{equation}
Symmetries~\eqref{permutation},~\eqref{exchange} together form the \textit{tetrahedral symmetry}. \\ \\
    3. Complex conjugation of quantum 6j-symbol is equivalent to conjugation of all representations:
    \begin{equation}
        \left\lbrace \begin{matrix}
	R_1 & R_2 & R_{3} \\
	r_1 & r_{2} & r_{3}
	\end{matrix} \right\rbrace^* = \left\lbrace \begin{matrix}
	\overline{R}_1 & \overline{R}_2 & \overline{R}_{3} \\
	\overline{r}_1 & \overline{r}_{2} & \overline{r}_{3}
	\end{matrix} \right\rbrace\,.
    \end{equation}
    4. There is the following unitarity property:
    \begin{equation}
        \sum\limits_{r_3} \text{qdim}(R_3)\,\text{qdim}(r_3)\left\lbrace \begin{matrix}
	R_1 & R_2 & R_{3} \\
	r_1 & r_{2} & r_{3}
	\end{matrix} \right\rbrace \left\lbrace \begin{matrix}
	R_1 & R_2 & R'_{3} \\
	r_1 & r_{2} & r_{3}
	\end{matrix} \right\rbrace^*=\delta_{R_3 R'_3}\,.
    \end{equation}
    5. The generalized Racah backcoupling rule:
    \begin{equation}
        q^{(\mathcal{C}_2^{R_1}+\mathcal{C}_2^{r_1}+\mathcal{C}_2^{R_3}+\mathcal{C}_2^{r_3})/2}\left\lbrace \begin{matrix}
	R_1 & R_2 & R_{3} \\
	r_1 & r_{2} & r_{3}
	\end{matrix} \right\rbrace = \sum\limits_{R} (-)^{\#} q^{(\mathcal{C}_2^R+\mathcal{C}_2^{R_2}+\mathcal{C}_2^{r_2})/2}\text{qdim}(R)\left\lbrace \begin{matrix}
	r_3 & R & R_{3} \\
	r_1 & r_{2} & R_{1}
	\end{matrix} \right\rbrace \left\lbrace \begin{matrix}
	R_1 & R_2 & R_{3} \\
	\overline{r}_3 & R & \overline{r}_{1}
	\end{matrix} \right\rbrace\,,
    \end{equation}
    where $\mathcal{C}_2^R$ is $\mathfrak{sl}_N$ quadratic Casimir for the representation $R$ and signs in the sum depend on all representations $R_{1,2,3}$, $r_{1,2,3}$ and $R$. \\ \\
6. The \textit{pentagon relation}:
    {\small \begin{equation}
        \left\lbrace \begin{matrix}
	R_1 & R_2 & R_{3} \\
	r_1 & r_{2} & r_{3}
	\end{matrix} \right\rbrace = \sum\limits_{R,R'_3}(-)^{\#}\text{qdim}(R_3)\,\text{qdim}(R'_3)\,\text{qdim}(R) \left\lbrace \begin{matrix}
	R'_2 & \overline{r}_2 & R \\
	r_3 & R'_{3} & \overline{R}_{1}
	\end{matrix} \right\rbrace \left\lbrace \begin{matrix}
	R'_3 & \overline{r}_3 & R \\
	r_1 & R'_{1} & \overline{R}_{2}
	\end{matrix} \right\rbrace \left\lbrace \begin{matrix}
	R'_1 & \overline{r}_1 & R \\
	r_2 & R'_{2} & \overline{R}_{3}
	\end{matrix} \right\rbrace \left\lbrace \begin{matrix}
	R_1 & R_2 & R_{3} \\
	R'_1 & R'_{2} & R'_{3}
	\end{matrix} \right\rbrace\,,
    \end{equation}}
    where signs in the sum depend on all representations $R_{1,2,3}$, $R'_{1,2,3}$, $r_{1,2,3}$ and $R$. For more accurate formulation of these symmetries see~\cite{lienert1992racah,Gu_2015}.

\subsection{$\mathcal{R}$-matrix}\label{R-matrix}
In this subsection, we define the second important object of our study -- the so-called \textit{quantum $\mathcal{R}$-matrix}. In WZW conformal field theory $\mathcal{R}$-matrices correspond to \textit{half-monodromy operators} acting on primary fields and are usually called \textit{braiding operators}.

Consider $U_q(\mathfrak{sl}_N)$ algebra irreducible finite-dimensional representations $V_{R_i}$, $i=1,\dots,m$.
\\ \\
\textit{$\mathcal{R}$-matrices are invertible linear operators defined by}
\begin{equation}\label{R-matrix}
\mathcal{R}_i=1_{V_{R_1}} \otimes 1_{V_{R_2}} \otimes \ldots \otimes P \check{\mathcal{R}}_{i, i+1} \otimes \ldots \otimes 1_{V_{R_m}} \in \operatorname{End}\left(V_{R_1} \otimes \ldots \otimes V_{R_m}\right)\,,
\end{equation}
\textit{where} $P(x\otimes y)=y\otimes x$ \textit{and} $\check{\mathcal{R}}$ \textit{is the universal $\mathcal{R}$-matrix}:
\begin{equation}\label{UniversalR}
\check{\mathcal{R}}=q^{\sum_{n, m} C_{n m}^{-1} H_n \otimes H_m} \prod_{\text {positive root } \beta} \exp _q\left[\left(q-q^{-1}\right) E_\beta \otimes F_\beta\right].
\end{equation}
\textit{Here $\exp_q$ is the quantum exponent, $C_{nm}$ is the Cartan matrix and $\{H_i,E_i,F_i\}$ are $U_q(\mathfrak{sl}_N)$ generators. $\mathcal{R}$-matrix $\check{\mathcal{R}}_{i,i+1}$ from~\eqref{R-matrix} acts on $V_{R_i}\otimes V_{R_{i+1}}$ and generators are taken in the corresponding representations $V_{R_i}$ and $V_{R_{i+1}}$.} \\ \\
It is well-known that $\mathcal{R}_i$, $i=1,\dots,m,$ define a representation of the Artin's braid group $\mathcal{B}_m$ on $m$ strands:
\begin{equation}
\pi: \mathcal{B}_m \rightarrow \operatorname{End}\left(V_{R_1} \otimes \ldots \otimes V_{R_m}\right)\,, \quad
\pi\left(\sigma_i\right) =\mathcal{R}_i\,,
\end{equation}
where $\sigma_1,\dots,\sigma_{m-1}$ are generators of the braid group $\mathcal{B}_m$. 

The eigenvalues of the universal $\mathcal{R}$-matrix $\check{\mathcal{R}}_{i, i+1}$ are well-known~\cite{Reshetikhin,gould1994quantum}:
\begin{equation}\label{R-eigen}
\begin{array}{l}
\lambda_{R_{i,i+1}}=\varepsilon_{R_{i,i+1}} q^{\varkappa\left(R_{i,i+1}\right)-\varkappa\left(R_i\right)-\varkappa\left(R_{i+1}\right)}, \quad \text { if } R_i \neq R_{i+1}\,,\\
\lambda_{R_{i,i+1}}=\varepsilon_{R_{i,i+1}} q^{\varkappa\left(R_{i,i+1}\right)-4 \varkappa\left(R_i\right)-\left|R_i\right| N}, \quad\;\, \text { if } R_i=R_{i+1}\,.
\end{array}
\end{equation}
Here index $i$ enumerates representations, and $V_{R_{i,i+1}}$ are irreducible components of the tensor product $V_{R_i}$ and $V_{R_{i+1}}\,$:
\begin{equation}\label{Irred}
V_{R_i} \otimes V_{R_{i+1}}=\bigoplus_{R_{i,i+1}} V_{R_{i,i+1}}\,,
\end{equation}
where we allow repeated summands; $\varkappa(R)$ is defined by the following formula:
\begin{equation}
\varkappa(R)=\sum_{(i, j) \in R} (i-j)\,,
\end{equation}
and $\epsilon_{R_{i,i+1}}=\pm 1$ is a sign. In the case when all representations are the same ($R_i=R$), sings of the eigenvalues depend on whether highest weight vectors of the representations are \textit{symmetric} or \textit{antisymmetric} under permutation of two representations $V_{R_i}$ and $V_{R_{i+1}}$. These two types of representations $V_{R_{i,i+1}}$ are said to belong to either symmetric, or antisymmetric squares of the representation $V_{R}$.

\subsection{$\mathcal{R}$-matrices via Racah matrices}\label{RviaRacah}
As we know the $\mathcal{R}$-matrix eigenvalues~\eqref{R-eigen}, we would like to diagonalize all $\mathcal{R}$-matrices. This can be done with the use of Racah matrices.

Consider $m$ representations $V_{R_i}$ and choose the basis in $V_{R_1} \otimes \cdots \otimes V_{R_m}$ in which the matrix $\mathcal{R}_1$ gets a block form. Such basis corresponds to the following order in the tensor product:
\begin{equation}
B_{12,3, . ., m}:=\left(\ldots\left(\left(V_{R_1} \otimes V_{R_2}\right) \otimes V_{R_3}\right) \otimes \ldots\right) \otimes V_{R_m}\,.
\end{equation}
Different blocks of $\mathcal{R}_1$ correspond to different representations $V_{R_{12}}$ in the decomposition
\begin{equation}
V_{R_1} \otimes V_{R_2}=\bigoplus_{R_{12}} \mathcal{M}_{R_{12}}^{R_1,R_2} \otimes V_{R_{12}}\,.
\end{equation}
If we additionally rotate the components corresponding to each $R_{12}$, we can diagonalize the matrix $\mathcal{R}_1$.

The same procedure can be made in the basis corresponding to
\begin{equation}
B_{1,23, . ., m}:=\left(\ldots\left(V_{R_1} \otimes\left(V_{R_2} \otimes V_{R_3}\right)\right) \otimes \ldots\right) \otimes V_{R_m}\,.
\end{equation}
Therefore, in order to diagonalize the matrix $\mathcal{R}_2$ one should make the basis transformation with the help of Racah matrix:
\begin{equation}
\mathcal{R}_2=U^{\dagger}\left[\begin{array}{ll}
R_1 & R_3 \\
R_2 & R_{123}
\end{array}\right] \cdot \Lambda_{\mathcal{R}_2} \cdot U\left[\begin{array}{ll}
R_1 & R_2 \\
R_3 & R_{123}
\end{array}\right]\,,
\end{equation}
where $\Lambda_{\mathcal{R}_2}$ is a diagonal matrix with eigenvalues of $\mathcal{R}_2$. The similar procedure can be made with all $\mathcal{R}$-matrices. While eigenvalues for all $\mathcal{R}$-matrices can be computed explicitly~\eqref{R-eigen}, calculation of Racah coefficients is a puzzling procedure, that was done analytically only for $U_q(\mathfrak{sl}_2)$ case.

\subsection{Eigenvalue hypothesis~\cite{itoyama2013eigenvalue}}\label{EigenHyp}
In this subsection, we introduce the eigenvalue hypothesis. Let us, first, describe some intuition on how the hypothesis can be formulated. The conjecture originates from the Yang-Baxter equation
\begin{equation}\label{YB}
    \mathcal{R}_1\mathcal{R}_2\mathcal{R}_1=\mathcal{R}_2\mathcal{R}_1\mathcal{R}_2\,.
\end{equation}
Consider the case when Racah matrix act in the tensor cube of a representation $V_R$ of $U_q(\mathfrak{sl}_N)$:
\begin{equation}\label{knotU}
    U:\quad (V_R\otimes V_R)\otimes V_R\rightarrow V_R\otimes (V_R\otimes V_R)\,.
\end{equation}
Let us choose the basis where $\mathcal{R}_1$ is diagonal. Then, following the pocedure described in Subsection~\ref{RviaRacah}, one can diagonalize $\mathcal{R}_2$ using Racah matrix: $\mathcal{R}_2=U^\dagger \mathcal{R}_1 U$, and the Yang-Baxter equation~\eqref{YB} takes the form
\begin{equation}\label{YB-mod}
    \mathcal{R}_1U^\dagger \mathcal{R}_1 U\mathcal{R}_1=U^\dagger \mathcal{R}_1 U\mathcal{R}_1U^\dagger \mathcal{R}_1 U\,.
\end{equation}
In this equation, $U$-matrix elements are expressed through $\mathcal{R}$-matrix eigenvalues. The solution of nonlinear equation~\eqref{YB-mod} is not unique. For example, there are obvious solutions $U=\pm\mathbb{I}$, where $\mathbb{I}$ is an identity matrix, but they correspond to $U$-matrix~\eqref{knotU} not for any representation $V_R$. Equation~\eqref{YB-mod} has two obvious symmetries. First, equation~\eqref{YB-mod} is invariant under multiplication of $\mathcal{R}_1$ by a constant. Second, this equation is invariant under a change of basis, which permutes $\mathcal{R}$-matrix eigenvalues. If for different representations $V_{R}$ and $V_{R'}$, $\mathcal{R}$-matrices eigenvalues differ by a common factor and by permutation (i.e. sets of normalized eigenvalues coincide), then equation~\eqref{YB-mod} remains the same. The eigenvalue hypothesis states that for arbitrary irreducible finite-dimensional $U_q(\mathfrak{sl}_N)$ representations Racah matrix $U\left[\begin{array}{ll}
R & R \\
R & Q
\end{array}\right]$ is \textit{uniquely} expressed through a set of normalized $\mathcal{R}$-matrix eigenvalues. The eigenvalue hypothesis also can be formulated in the case of three different irreducible $U_q(\mathfrak{sl}_N)$ representations as follows. \\ \\
\textbf{Eigenvalue conjecture~\cite{itoyama2013eigenvalue}.} \textit{Given arbitrary irreducible finite-dimensional $U_q(\mathfrak{sl}_N)$ representations for generic $q$, Racah matrix $U\left[\begin{array}{ll}
R_1 & R_2 \\
R_3 & R_{123}
\end{array}\right]$ is uniquely expressed through 3 sets of normalized eigenvalues of $\check{\mathcal{R}}_{1,2}$, $\check{\mathcal{R}}_{2,3}$ and $\check{\mathcal{R}}_{1,3}$ defined in~\eqref{UniversalR}.}
\\ \\
As a straightforward corollary, we get that if one finds a transformation of representations which does not change quantum $\mathcal{R}$-matrices eigenvalues, it correspond to a symmetry of quantum Racah coefficients with respect to the same transformation of representations.

\subsection{Colored HOMFLY polynomial}\label{HOMFLYdefSec}
In this subsection, we define the colored HOMFLY polynomial with the use of Reshetikhin-Turaev approach~\cite{reshetikhin1990ribbon}. \\ \\
\textbf{Alexander theorem.} \textit{Any link $\mathcal{L}$ in $\mathbb{R}^3$ can be obtained by a closure of the corresponding braid.}\\ \\
If a link $\mathcal{L}$ has one component, it is called a \textit{knot} and is usually denoted $\mathcal{K}$. Each link component must carry its own representation. In particular, for a knot each strand of the corresponding braid carries one and the same representation.

Let $\mathcal{L}$ be an oriented link with $L$ components $\mathcal{K}_1,\dots,\mathcal{K}_L$ colored by irreducible finite-dimensional representations $V_{R_1},\dots,V_{R_L}$ of $U_q(\mathfrak{sl}_N)$, and $\beta_L\in\mathcal{B}_m$ is some $m$-strand braid which closure gives $\mathcal{L}$.\\ \\
\textit{The colored HOMFLY polynomial is a quantum group invariant of a link $\mathcal{L}$ defined as follows\footnote{In order to provide the invariance under the first Reidemeister move, one puts the usual framing factor~\cite{liu2010proof} in front of the quantum trace. We incorporate it in $\mathcal{R}$-matrix by modifying its eigenvalues~\eqref{R-eigen}.}:}
\begin{equation}\label{HOMFLYdef}
H_{R_1, \ldots, R_L}^{\mathcal{L}}={ }_q \operatorname{tr}_{V_{R_1} \otimes \cdots \otimes V_{R_m}}\left(\pi\left(\beta_{\mathcal{L}}\right)\right),
\end{equation}
\textit{where ${ }_q \operatorname{tr}$ is the quantum trace.} \\ \\
One can expand the quantum trace in~\eqref{HOMFLYdef}:
\begin{equation}\label{R-THOMFLY}
    H_{R_1, \ldots, R_L}^{\mathcal{L}}(q,A=q^N)=\sum\limits_{V_Q\in V_{R_1} \otimes \cdots \otimes V_{R_m}} \operatorname{tr}_{\mathcal{M}_{Q}}\left(\pi\left(\beta_{\mathcal{L}}\right)\right) \cdot \mathrm{qdim}(Q),
\end{equation}
where we decompose $\bigotimes\limits_{i=1}^m V_{R_i}=\bigoplus_Q \mathcal{M}_{Q} \otimes V_Q\,.$

It was a breakthrough when the connection with topological field theory was established~\cite{Witten}. Namely, the colored HOMFLY polynomial turned out to be the Wilson loop in 3d Chern-Simons theory with $SU(N)$ gauge group defined by representation $R$ and knot $\mathcal{K}\,$:
     \begin{equation}
     \label{WilsonLoopExpValue}
         \mathcal{H}_{R}^{\mathcal{K}}(q, a) = \frac{1}{\text{qdim}(R)}\left\langle \text{tr}_{R} \ \text{Pexp} \left( \oint_{\mathcal{K}} A \right) \right\rangle_{\text{CS}},
     \end{equation}
     where Pexp denotes a path-ordered exponential, $\frac{1}{\text{qdim}(R)}$ is the normalization factor and the Chern-Simons action is given by
     \begin{equation}\label{CSAction}
         S_{\text{CS}}[A] = \frac{\kappa}{4 \pi} \int_{S^3} \text{tr} \left(  A \wedge dA +  \frac{2}{3} A \wedge A \wedge A \right).
     \end{equation}
     For links, the colored HOMFLY polynomial is defined by several Wilson loop operators, each one is associated with separate link component and carrying its own representation.

\setcounter{equation}{0}
\section{Tug-the-hook symmetry for Racah coefficients}\label{TTH}
Symmetries of 6j-symbols have been completely described for $U_q(\mathfrak{sl}_2)$ case only~\cite{kirillow1989representations}. The corresponding symmetry group is $S_4\times S_3$, where $S_4$ is responsible for the tetrahedral symmetry, and $S_3$ stands for the Regge symmetry. For $U_q(\mathfrak{sl}_N)$, $N>2$, the full symmetry group is unknown. However, there is a work~\cite{AMorozov_2018} on the possibility of construction of new symmetries with the use of the eigenvalue hypothesis, but explicit symmetries were provided only for the cases of symmetric and their conjugate representations in~\cite{Alekseev_2020,Alekseev_2021}. 

In this section, we introduce the tug-the-hook symmetry formulated in terms of Young diagrams transformation which leaves Racah coefficients invariant. This symmetry has a significant property. It transforms \textit{any} Young diagram which can be placed inside a hook, including cases with multiplicities. This is the first found symmetry of Racah matrices which is valid beyond multiplicity-free cases.

Let us introduce the tug-the-hook transformation~\cite{MST2} in terms of Young diagrams. A Young diagram is placed inside an appropriate $(K+M|M)$ fat hook for some integers $K$ and $M$. Introduce an analogue of Frobenius notations: parametrize the first $K$ rows by their length $R_i$, $i=1,\dots,K$, the rest rows are parametrized by shifted Frobenius variables $\alpha_i =R_i-(i-K)+1$, $\beta_i =R_{i-K}^{\prime}-i+1$, $i=K+1,\dots,K+M$. The tug-the-hook transformation $\mathbf{T}_\epsilon^{(K+M|M)}$ pulls the Young diagram inside the fat hook:
\begin{equation}
R_i \longrightarrow R_i-\epsilon\,, \quad
\alpha_i \longrightarrow \alpha_i-\epsilon\,, \quad
\beta_i \longrightarrow \beta_i+\epsilon\,,
\end{equation}
where $\epsilon$ is the corresponding shift of the diagram. A shift $\epsilon$ can be negative, what corresponds to the inverse shift. Note that this transformation is defined when the resulting shifted Frobenius variables form a Young diagram. An example is shown in Fig.~\ref{TTHpic}.
\tikzfading[name=fade right,
left color=transparent!98,
right color=transparent!95]
\tikzfading[name=fade down,
top color=transparent!98,
bottom color=transparent!95]
\begin{center}
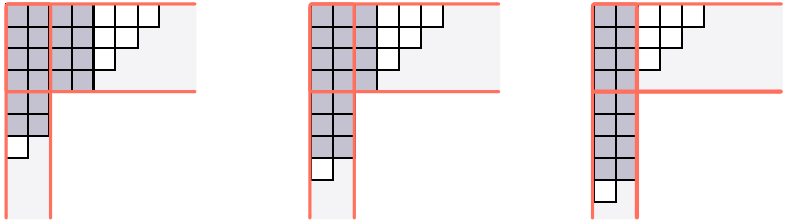
\begin{figure}[h!]
\centering
\resizebox{0.6\textwidth}{!}{%
\begin{tikzpicture}[>=triangle 45,font=\sffamily,rounded corners=1pt]
    \node  (1) at (0,0) {$\ytableausetup
{boxsize=1em}
\ytableausetup{aligntableaux = top}
\ydiagram[*(white)]{4+3,4+2,4+1,0,0,0,1}*[*(lavendergray)]{4,4,4,4,2,2} \hspace{2.5cm} \ydiagram[*(white)]{3+3,3+2,3+1,0,0,0,0,1}*[*(lavendergray)]{3,3,3,3,2,2,2} \hspace{2.5cm}  \ydiagram[*(white)]{2+3,2+2,2+1,0,0,0,0,0,1}*[*(lavendergray)]{2,2,2,2,2,2,2,2}$};

\draw[line width=0.15em,bittersweet,round cap-round cap]([xshift=0.4em,yshift=-0.45em]1.north west)--([xshift=0.4em,yshift=-10.55em]1.north west);
\draw[line width=0.15em,bittersweet,round cap-round cap]([xshift=2.5em,yshift=-0.45em]1.north west)--([xshift=2.5em,yshift=-10.55em]1.north west);
\draw[line width=0.15em,bittersweet,round cap-round cap]([xshift=0.4em,yshift=-0.35em]1.north west)--([xshift=9.4em,yshift=-0.35em]1.north west);
\draw[line width=0.15em,bittersweet,round cap-round cap]([xshift=0.4em,yshift=-4.5em]1.north west)--([xshift=9.4em,yshift=-4.5em]1.north west);
\draw [line width=0.15em,bittersweet]([xshift=0.4em,yshift=-0.35em]1.north west) rectangle ([xshift=2.5em,yshift=-4.5em]1.north west);
\begin{scope}[on background layer]
\fill [anti-flashwhite,opacity=0.8]([xshift=0.4em,yshift=-0.35em]1.north west) rectangle ([xshift=2.5em,yshift=-10.55em]1.north west);
\fill [anti-flashwhite,opacity=0.8]([xshift=0.4em,yshift=-0.35em]1.north west) rectangle ([xshift=9.4em,yshift=-4.5em]1.north west);
\end{scope}
\draw[line width=0.15em,bittersweet,round cap-round cap]([xshift=0.4em+7.25em+2.5cm,yshift=-0.45em]1.north west)--([xshift=0.4em+7.25em+2.5cm,yshift=-10.55em]1.north west);
\draw[line width=0.15em,bittersweet,round cap-round cap]([xshift=2.5em+7.25em+2.5cm,yshift=-0.45em]1.north west)--([xshift=2.5em+7.25em+2.5cm,yshift=-10.55em]1.north west);
\draw[line width=0.15em,bittersweet,round cap-round cap]([xshift=0.4em+7.25em+2.5cm,yshift=-0.35em]1.north west)--([xshift=9.4em+7.25em+2.5cm,yshift=-0.35em]1.north west);
\draw[line width=0.15em,bittersweet,round cap-round cap]([xshift=0.4em+7.25em+2.5cm,yshift=-4.5em]1.north west)--([xshift=9.4em+7.25em+2.5cm,yshift=-4.5em]1.north west);
\draw [ line width=0.15em,bittersweet]([xshift=7.65em+2.5cm,yshift=-0.35em]1.north west) rectangle ([xshift=7.6em+2.5em-0.35em+2.5cm,yshift=-4.5em]1.north west);
\begin{scope}[on background layer]
\fill [anti-flashwhite,opacity=0.8]([xshift=0.4em+7.25em+2.5cm,yshift=-0.35em]1.north west) rectangle ([xshift=2.5em+7.25em+2.5cm,yshift=-10.55em]1.north west);
\fill [anti-flashwhite,opacity=0.8]([xshift=0.4em+7.25em+2.5cm,yshift=-0.35em]1.north west) rectangle ([xshift=9.4em+7.25em+2.5cm,yshift=-4.5em]1.north west);
\end{scope}
\draw[line width=0.15em,bittersweet,round cap-round cap]([xshift=0.4em+13.5em+5cm,yshift=-0.45em]1.north west)--([xshift=0.4em+13.5em+5cm,yshift=-10.55em]1.north west);
\draw[line width=0.2em,bittersweet,round cap-round cap]([xshift=2.5em+13.5em+5cm,yshift=-0.45em]1.north west)--([xshift=2.5em+13.5em+5cm,yshift=-10.55em]1.north west);
\draw[line width=0.15em,bittersweet,round cap-round cap]([xshift=0.4em+13.5em+5cm,yshift=-0.35em]1.north west)--([xshift=9.4em+13.5em+5cm,yshift=-0.35em]1.north west);
\draw[line width=0.2em,bittersweet,round cap-round cap]([xshift=0.4em+13.5em+5cm,yshift=-4.5em]1.north west)--([xshift=9.4em+13.5em+5cm,yshift=-4.5em]1.north west);
\draw [ line width=0.15em,bittersweet]([xshift=13.9em+5cm,yshift=-0.35em]1.north west) rectangle ([xshift=13.9em+2.5em-0.4em+5cm,yshift=-4.5em]1.north west);
\begin{scope}[on background layer]
\fill [anti-flashwhite,opacity=0.8]([xshift=0.4em+13.5em+5cm,yshift=-0.35em]1.north west) rectangle ([xshift=2.5em+13.5em+5cm,yshift=-10.55em]1.north west);
\fill [anti-flashwhite,opacity=0.8]([xshift=0.4em+13.5em+5cm,yshift=-0.35em]1.north west) rectangle ([xshift=9.4em+13.5em+5cm,yshift=-4.5em]1.north west);
\end{scope}
\end{tikzpicture}
 }%
\caption{Example of applying the tug-the-hook transformations $\mathbf{T}_1^{(4|2)}$ from the left diagram to the middle one and $\mathbf{T}_2^{(4|2)}$ from the left diagram to the right one~\cite{MST2}.}\label{TTHpic}
\end{figure}
\end{center}
\vspace{-0.8cm}
We put forward the following conjecture.\\ \\
 \textit{Given arbitrary irreducible finite-dimensional representations $V_{R_1}$, $V_{R_2}$, $V_{R_3}$, $V_{R_{12}}\in V_{R_1}\otimes V_{R_2}$, $V_{R_{23}}\in V_{R_2}\otimes V_{R_3}$, $V_{R_{123}}\in V_{R_1}\otimes V_{R_2}\otimes V_{R_3}$ of $U_q(\mathfrak{sl}_N)$ for generic $q$, Racah coefficients are invariant under the tug-the-hook transformation:}
\begin{equation}
    U\left[\begin{array}{cc|c} R_1 & R_2 & R_{12} \\ R_3 & R_{123} & R_{23} \end{array}\right]=U\left[\begin{array}{cc|c} \mathbf{T}^{(K+M|M)}_{\epsilon_1}(R_1)&\mathbf{T}^{(K+M|M)}_{\epsilon_2}(R_2)& \mathbf{T}^{(K+M|M)}_{\epsilon_1+\epsilon_2}(R_{12}) \\[2.5pt] \mathbf{T}^{(K+M|M)}_{\epsilon_3}(R_3)&\mathbf{T}^{(K+M|M)}_{\epsilon_1+\epsilon_2+\epsilon_3}(R_{123}) &\mathbf{T}^{ (K+M|M)}_{\epsilon_2+\epsilon_3}(R_{23})\end{array}\right],
\end{equation}
\textit{for $K$, $M$ and integer shifts $\epsilon_1$, $\epsilon_2$, $\epsilon_3$ for which the tug-the-hook transformation is defined.} \\ \\
This hypothesis follows from the eigenvalue conjecture. The \textit{proof} consists of the following steps.
\begin{enumerate}
    \item $\mathfrak{sl}_N$ Casimir invariants are invariant under the tug-the-hook transformation what was proved in our recent paper~\cite{Lanina:2021nfj}.
    \item $\mathcal{R}$-matrices eigenvalues~\eqref{R-eigen} can be expressed through $\mathfrak{sl}_N$ quadratic Casimir invariants $\mathcal{C}_2$~\cite{Reshetikhin,klimyk2012quantum}:
    \begin{equation}
        \lambda_Q=\varepsilon_Q\, q^{(\mathcal{C}^Q_2-\mathcal{C}^{R}_2-\mathcal{C}^{R'}_2)/2}\,,\quad Q\in R\otimes R'\,,
    \end{equation}
    and signs $\varepsilon_Q$ for all $Q$ are multiplied by one and the same factor $\pm 1$ under the tug-the-hook transformation, what does not change the normalized eigenvalues. Thus, normalized $\mathcal{R}$-matrices eigenvalues are also invariant under the tug-the-hook transformation.
    \item Thus, the eigenvalue hypothesis implies that Racah coefficients possess the tug-the-hook symmetry. 
\end{enumerate} 
Emphasise several important peculiarities of the tug-the-hook symmetry for quantum Racah coefficients. 
\\ \\
\textbf{Remark 1.} \textit{Due to the fact that Racah coefficients do not depend on $K$ and $M$, the tug-the-hook symmetry holds for any hook which the corresponding Young diagrams can fit.} \\ \\
\textbf{Remark 2.} \textit{The tug-the-hook symmetry was first observed~\cite{MST2} and then proved~\cite{Lanina:2021nfj} for the colored HOMFLY polynomials for knots. In that case, the symmetry explicitly depends on $\mathfrak{sl}_{N+1}$ algebra rank $N$ because it acts only inside fat hooks $(N+M|M)$. This fact also means that lenghts of Young diagrams must be larger than $N$, and the tug-the-hook symmetry for the HOMFLY polynomials has $\mathfrak{sl}_{N+M|M}$ supergroup origins, where it is well-defined for its finite-dimensional representations. Unlike the HOMFLY case, the tug-the-hook symmetry for Racah coefficients is independent on $N$ and holds for corresponding representations of any $\mathfrak{sl}_{N+1}$ algebra.}\\ \\
\textbf{Remark 3.} \textit{The tug-the-hook symmetry is the first found symmetry of Racah matrices which acts on any representations and works for cases with multiplicities.}
\\ \\
The simplest non-trivial example is the case without multiplicities and $2\times 2$ matrices. Using database~\cite{knotebook} and explicit formulas for Racah matrices for symmetric incoming representations~\cite{Dhara_2019}, we immediately see that
\begin{equation}
\begin{aligned}
    &U\begin{bmatrix} [2,2] & [2,2] \\ [2,2] & [5,4,1,1,1]
    \end{bmatrix}=U\begin{bmatrix} \mathbf{T}_1^{(2|1)}([2,2]) & \mathbf{T}_1^{(2|1)}([2,2]) \\ \mathbf{T}_1^{(2|1)}([2,2]) & \mathbf{T}_3^{(2|1)}([5,4,1,1,1])
    \end{bmatrix}=U\begin{bmatrix} [3]' & [3]' \\ [3]' & [8,1]'
    \end{bmatrix}=U\begin{bmatrix} [3] & [3] \\ [3] & [8,1]
    \end{bmatrix},\\
    &U\begin{bmatrix} [2,2] & [2,2] \\ [2,2] & [5,4,1,1,1]
    \end{bmatrix}=U\begin{bmatrix} \mathbf{T}_{-1}^{(1|1)}([3]) & \mathbf{T}_{-1}^{(1|1)}([3]) \\ \mathbf{T}_{-1}^{(1|1)}([3]) & \mathbf{T}_{-3}^{(1|1)}([8,1])
    \end{bmatrix}=U\begin{bmatrix} [2,1] & [2,1] \\ [2,1] & [5,1,1,1,1]
    \end{bmatrix}.
\end{aligned}
\end{equation}

\setcounter{equation}{0}
\section{Evidence for the tug-the-hook symmetry}
In this section, we provide facts and examples of the validity of conjecture that quantum 6j-symbols possess the tug-the-hook symmetry. Unlike the example of the previous section, here we consider cases with multiplicities where the eigenvalue conjecture is not proved. Still, in these cases the tug-the-hook symmetry for Racah coefficients holds. This fact also indirectly confirms the eigenvalue hypothesis. 

\subsection{Proofs for the eigenvalue hypothesis}\label{EigHypProofs}
The eigenvalue hypothesis has been proved for several specific cases:
\begin{itemize}
    \item for $U_q(\mathfrak{sl}_2)$ in~\cite{Alekseev_2021};
    \item in multiplicity-free $U_q(\mathfrak{sl}_N)$ case for coinciding incoming representations for matrices up to size $5\times 5$~\cite{arxiv.math/9912013,itoyama2013eigenvalue}.
\end{itemize}
Thus, for these cases, the tug-the-hook symmetry for Racah matrices is also proved.

\subsection{Tug-the-hook symmetry for the colored HOMFLY polynomials}\label{HOMFLY}
In recent paper~\cite{Lanina:2021nfj}, the tug-the-hook symmetry for the HOMFLY polynomials for knots has been proved. It is of physical importance, as it connects Wilson loops in the 3d Chern-Simons theory~\eqref{WilsonLoopExpValue} in different representations. In that case the tug-the-hook symmetry is specified: $K=N$, where $N$ is $\mathfrak{sl}_{N+1}$ algebra rank:
\begin{equation}
\mathcal{H}_R^{\mathcal{K}}\left(q, A=q^N\right)=\mathcal{H}^{\mathcal{K}}_{\mathbf{T}_\epsilon^{(N+M|M)}(R)}\left(q, A=q^N\right)\,.
\end{equation}
Let us examine a consequence of this symmetry. First, note that
\begin{equation}
\frac{\operatorname{qdim}\left(\mathbf{T}_\epsilon^N(R)\right)}{\operatorname{qdim}(R)}=(-1)^{\epsilon h_N(R)},
\end{equation}
where $h_N(R)$ is the number of boxes on the shifted downwards by $N$ diagonal of the Young diagram $R$. The \textit{normalized} HOMFLY polynomials are constructed from traces of products of $\mathcal{R}$-matrices and quantum dimensions~\eqref{R-THOMFLY}:
\begin{equation}
\mathcal{H}_R^{\mathcal{K}}=\sum_{V_Q \in V_R^{\otimes m}} \operatorname{tr}_{\mathcal{M}_{Q}}\left(\pi\left(\beta_{\mathcal{K}}\right)\right) \frac{\mathrm{qdim}(Q)}{\mathrm{qdim}(R)}\,.
\end{equation}
Thus, one expects that $\operatorname{tr}_{\mathcal{M}_{Q}}\left(\pi\left(\beta_{\mathcal{K}}\right)\right)$ are preserved (up to the sign) under the tug-the-hook transformation too.

For clearness, consider an example of the 3-strand braid, for which the HOMFLY polynomial
\begin{equation}
    \mathcal{H}_R^{\mathcal{K}}=\sum_{V_Q \in V_R^{\otimes 3}} \operatorname{tr}_{\mathcal{M}_{Q}}\left(\prod_{i=1}^n\mathcal{R}^{a_i}_1\mathcal{R}^{b_i}_2\right) \frac{\mathrm{qdim}(Q)}{\mathrm{qdim}(R)}
\end{equation}
with arbitrary amount $n$ of $a_i$, $b_i\in\mathbb{Z}$. As it is described in Subsection~\ref{RviaRacah}, we choose the basis in which $\mathcal{R}_1$ is diagonal $\mathcal{R}_1=\Lambda_{\mathcal{R}_1}$. Then, $\mathcal{R}_2$ is diagonalized by the corresponding $U$-matrix: $\mathcal{R}_2=U^\dagger \Lambda_{\mathcal{R}_1} U$. $\Lambda_{\mathcal{R}_1}$ is preserved under the tug-the-hook transformation, and $\operatorname{tr}_{\mathcal{M}_{Q}}\left(\prod_{i=1}^n\Lambda^{a_i}_{\mathcal{R}_1}U^\dagger \Lambda^{b_i}_{\mathcal{R}_1} U\right)$ should be invariant under this transformation for any $n$, $a_i$, $b_i$, $i=1,\dots,n$. So, it is naturally to expect that Racah coefficients possess the tug-the-hook symmetry when all representations in the tensor product are equal.

From the other hand, as we have already noted, for the general case of different incoming representations, $\mathcal{R}$-matrices eigenvalues conserve under the tug-the-hook transformation too. Moreover, our conjecture still holds in this general case, so Racah matrices possess the tug-the-hook symmetry. Thus, all building blocks of the link HOMFLY polynomial are invariant under the tug-the-hook transformation, and we can state that \textit{the colored HOMFLY polynomial for links possess the tug-the-hook symmetry.} Note that the tug-the-hook symmetry for the HOMFLY polynomials for links was neither proved, nor observed, and here we firstly claim it.

\subsection{$U_q(\mathfrak{sl}_N)$ examples}\label{TTHEx}
In this subsection, we give a list of highly nontrivial examples of the tug-the-hook symmetry for Racah matrices $U\left[\begin{array}{ll}
R & R \\
R & R_{123}
\end{array}\right]$ for cases where the eigenvalue hypothesis is still not proved: for cases with multiplicities. From the whole known 6j-symbols~\cite{knotebook}, the only one case turns out to be interesting -- the case $R=[3,2]\rightarrow\mathbf{T}_{1}^{(2|1)}([3,2])=[2,1,1]=[3,1]'$. Among all the resulting Young diagrams $R_{123}$, there are four nontrivial matrices with multiplicities, which give 99 6j-symbols in total.
\begin{enumerate}
    \item $V_{[7,5,1,1,1]}\in V_{[3,2]}^{\otimes 3}\rightarrow V_{[8,2,1,1]'}\in V_{[3,1]'}^{\otimes 3}$
\end{enumerate}
Consider this case in more details. An important peculiarity of Racah matrices computation~\cite{knotebook} is the fact that they were calculated in a specific basis of highest weight vectors. The bases of different Racah matrices are in correspondence if there is no multiplicities, because there is only one vector related to the corresponding highest weight. When multiplicities occur, the corresponding basis elements can be differently mixed. Thus, in order to show the validity of the tug-the-hook symmetry for Racah matrices, one should make an orthogonal transformation. In our case, there are two $6\times 6$ matrices with multiplicities: $U\begin{bmatrix}
[3,2] & [3,2] \\
[3,2] & [7,5,1,1,1]
\end{bmatrix}$ and $U\begin{bmatrix}
[3,1] & [3,1] \\
[3,1] & [8,2,1,1]
\end{bmatrix}$. The basis for $U\begin{bmatrix}
[3,2] & [3,2] \\
[3,2] & [7,5,1,1,1]
\end{bmatrix}$ from~\cite{knotebook} corresponds to the Young diagrams $[6,3,1]_-$, $[5,4,1]_+$, $[5,4,1]_-$, $2[5,3,1,1]_+$, $[4,4,1,1]_-$ arranged in the written lexicographical order, where $+$ index signals that the corresponding representation belongs to the symmetric square of $V_{[3,2]}$ and $-$ index signals that the corresponding representation belongs to the antisymmetric square of $V_{[3,2]}$. Applying the tug-the-hook transformation and transposing the diagrams, we arrive to $[5,1,1,1]_+$, $[5,2,1]_-$, $[5,2,1]_+$, $2[6,1,1]_-$, $[6,2]_+$ respectively, which should be arranged in the reverse order to form the basis for $U\begin{bmatrix}
[3,1] & [3,1] \\
[3,1] & [8,2,1,1]
\end{bmatrix}$ from~\cite{knotebook}. Fortunately, one needs to rotate only one block corresponding to $2[6,1,1]_-\,$:
\begin{equation}
    \widetilde{U}\begin{bmatrix}
[3,1] & [3,1] \\
[3,1] & [8,2,1,1]
\end{bmatrix}=\widetilde{O}'\,U\begin{bmatrix}
[3,1] & [3,1] \\
[3,1] & [8,2,1,1]
\end{bmatrix}O\,,
\end{equation}
where
\begin{equation}
    O=\left(
\begin{array}{cccccc}
 1 & 0 & 0 & 0 & 0 & 0 \\
 0 & \cos \theta & \sin \theta & 0 & 0 & 0 \\
 0 & -\sin \theta & \cos \theta & 0 & 0 & 0 \\
 0 & 0 & 0 & 1 & 0 & 0 \\
 0 & 0 & 0 & 0 & 1 & 0 \\
 0 & 0 & 0 & 0 & 0 & 1 \\
\end{array}
\right),\quad \widetilde{O}'=\left(
\begin{array}{cccccc}
 1 & 0 & 0 & 0 & 0 & 0 \\
 0 & \cos \tilde{\theta} & -\sin \tilde{\theta} & 0 & 0 & 0 \\
 0 & \sin \tilde{\theta} & \cos \tilde{\theta} & 0 & 0 & 0 \\
 0 & 0 & 0 & 1 & 0 & 0 \\
 0 & 0 & 0 & 0 & 1 & 0 \\
 0 & 0 & 0 & 0 & 0 & 1 \\
\end{array}
\right).
\end{equation}
There are unique (up to $2\pi k$, $k\in\mathbb{Z}$) angles $\theta$ and $\tilde{\theta}$ which results to
\begin{equation}\label{U=UTTH}
    \widetilde{U}\begin{bmatrix}
[3,1] & [3,1] \\
[3,1] & [8,2,1,1]
\end{bmatrix}=U\begin{bmatrix}
[3,2] & [3,2] \\
[3,2] & [7,5,1,1,1]
\end{bmatrix}.
\end{equation}
The answers are
\begin{equation}
\begin{aligned}
 \tan\theta&=-\frac{q^{11}}{\sqrt{q^4-q^3+q^2-q+1} \sqrt{q^4+q^3+q^2+q+1} \left(q^8+1\right)}\times\\&\times\frac{1}{ \sqrt{q^{10}-q^9+q^8-q^7+q^6-q^5+q^4-q^3+q^2-q+1} \sqrt{q^{10}+q^9+q^8+q^7+q^6+q^5+q^4+q^3+q^2+q+1}}\,, \\
  \tan\tilde{\theta}&=q \sqrt{q^4-q^3+q^2-q+1} \sqrt{q^4+q^3+q^2+q+1}\times\\&\times\frac{ \sqrt{q^{10}-q^9+q^8-q^7+q^6-q^5+q^4-q^3+q^2-q+1} \sqrt{q^{10}+q^9+q^8+q^7+q^6+q^5+q^4+q^3+q^2+q+1}}{\left(q^2-q+1\right) \left(q^2+q+1\right) \left(q^6-q^3+1\right) \left(q^6+q^3+1\right)}\,,
\end{aligned}
\end{equation}
and equality~\eqref{U=UTTH} holds for all matrix elements. In order to obtain the final coincidence, we transpose the Young diagrams, what does not change the $U$-matrix~\eqref{UT}.

Performing similar procedures, we also check the invariance of Racah matrices under the following tug-the-hook transformations:
\begin{enumerate}
    \item[2.] $V_{[8,5,1,1]}\in V_{[3,2]}^{\otimes 3}\rightarrow V_{[7,2,1,1,1]'}\in V_{[3,1]'}^{\otimes 3}$
    \item[3.] $V_{[8,6,1]}\in V_{[3,2]}^{\otimes 3}\rightarrow V_{[6,2,2,1,1]'}\in V_{[3,1]'}^{\otimes 3}$
    \item[4.] $V_{[7,6,1,1]}\in V_{[3,2]}^{\otimes 3}\rightarrow V_{[7,2,2,1]'}\in V_{[3,1]'}^{\otimes 3}$
\end{enumerate}
These examples provide very important checks for the tug-the-hook symmetry for Racah matrices.

\setcounter{equation}{0}
\section{Conclusion}
In this paper, we have introduced a new symmetry of quantum 6j-symbols. It is a very important result, as this symmetry can be applied to any representations which fit a corresponding hook, and holds even in cases with multiplicities. Previously known symmetries concerned only $U_q(\mathfrak{sl}_2)$ case and $U_q(\mathfrak{sl}_N)$, $N>2$, for symmetric representations and their conjugate ones. 

The tug-the-hook symmetry for 6j-symbols follows from the eigenvalue hypothesis. However, we have provided evidence for this symmetry which do not rely on the eigenvalue conjecture. First, we have explicitly checked that for the known Racah matrices the symmetry actually holds. Second, the tug-the-hook symmetry was proved for the colored HOMFLY polynomials for knots, and the tug-the-hook symmetry for Racah matrices for equal incoming representations should follow from the one for the HOMFLY polynomials, as Racah matrix is one of building blocks of the HOMFLY polynomial. From the other hand, using the eigenvalue conjecture, one can derive the tug-the-hook symmetry for the HOMFLY polynomials for links.

This work also demonstrates the importance of the eigenvalue conjecture. Namely, one can find new symmetries of 6j-symbols and quantum knot invariants by observing invariance of $\mathcal{R}$-matrices eigenvalues.

\setcounter{equation}{0}
\section*{Acknowledgements}


This work was funded by the grants of the Foundation for the Advancement of Theoretical Physics “BASIS” (E.L. and A.S.) and by RFBR grants 20-01-00644-A
(A.S.), 21-51-46010-ST-a (A.S.). We would like to thank Nikita Tselousov, Victor Mishnyakov and Victor Alexeev for useful discussions. 



\printbibliography

\end{document}